\documentclass[12pt]{iopart}
\usepackage{times}
\usepackage{epsfig}
\usepackage{graphicx}
\usepackage{dcolumn}
\usepackage{natbib}

\def\beginwide{
        \end{multicols} \vspace*{-0.5cm} \noindent
        \rule{3.5in}{.1mm}\rule{.1mm}{5mm} \widetext \medskip }
\def\beginwidetop{
        \end{multicols} \vspace*{-0.5cm} \noindent
        \widetext \medskip }
\def\endwide{
        \hspace*{3.35in}~\rule[-5mm]{.1mm}{5mm}\rule{3.5in}{.1mm}
        \begin{multicols}{2} \vspace*{-1.0cm} \noindent }
\def\endwidebottom{
        \begin{multicols}{2} \vspace*{-1.0cm} \noindent }

\begin{document}
\title{Power spectra of self-organized critical sandpiles}
\author{Lasse Laurson$^1$, Mikko J. Alava$^1$ and Stefano Zapperi$^{2,3}$}
\address{$^1$ Helsinki University of Technology, Laboratory of Physics,  
HUT-02105 Finland}
\address{$^2$ CNR-INFM, SMC, Dipartimento di Fisica,
Universit\`a "La Sapienza", P.le A. Moro 2
00185 Roma, Italy}
\address{$^3$CNR, Istituto dei Sistemi Complessi, Roma, Italy}
\begin{abstract}
We analyze the power spectra of avalanches in two classes of self-organized 
critical sandpile models, the Bak-Tang-Wiesenfeld model and the Manna model. 
We show that these decay with a
$1/f^\alpha$ power law, where the exponent value $\alpha$ is significantly smaller than $2$ and equals
the scaling exponent relating the avalanche size to its duration. 
We discuss the basic ingredients behind this result, such as the scaling of the
average avalanche shape.
\end{abstract}
\maketitle

Sandpile models have been introduced almost twenty years ago as a
paradigmatic example of self-organized criticality (SOC) \cite{BAK-87}, 
the tendency
of slowly driven dissipative systems to display a scale free
avalanche response.  Such ideas have had an enormous impact in different
fields, ranging from magnetic systems \cite{DUR-05}, 
superconductors \cite{FIE-95}, mechanics 
\cite{PET-94,SAL-02}, to geophysics and plasma physics including
in particular the magnetosphere \cite{reviews,journals,magnetos}.  
The influence also extends beyond physics, to for example
biology \cite{bio}, human (heart) physiology \cite{heart}, and
cognitive processes or neuroscience \cite{cogn}.

The reason behind this success lies in
the wide variety of non-equilibrium systems displaying an avalanche
response to an external driving.  
One of the primary aims of SOC was, originally, to explain the wide
occurrence of $1/f^\alpha$ noise in natural phenomena, through a direct
relation between avalanche scaling and spectral properties
\cite{BAK-87}. This idea was soon refuted when two groups
\cite{JEN-89,KER-90} published works independently claiming
that sandpile models should
lead instead to a Lorentzian spectrum, that is decaying as $1/f^2$ at large
frequencies. The theoretical arguments were supported by numerical
simulations on relatively small system sizes \cite{JEN-89,KER-90}. 
A non-trivial $1/f^\alpha$-decay in power spectra has only been 
found in non-critical sandpiles \cite{1/f}, or by an
alternative definition of the noise signal \cite{1/f2} , 
but not in standard cases such as in the 
original Bak-Tang-Wiesenfeld (BTW) model \cite{BAK-87}
and stochastic Manna model \cite{MAN-91}.

Sandpile models represent a useful idealization of avalanche
propagation, capturing the main ingredients behind this process: a slow
external driving, a local threshold - or non-linearity -
for the dynamics and a dissipation
mechanism.  While a complete exact solution of sandpile models is
possible only in some particular cases \cite{DHA-99}, the origin of the 
scaling behavior is now well understood in the realm of non-equilibrium 
critical phenomena \cite{DIC-00,VES-97,DIC-98}.  Systems presenting a
transition from an absorbing state to a moving phase, or similarly a
depinning transition \cite{ALA-01,ALA-04}, can be turned into SOC under a
suitable combination of a driving and a dissipation mechanism
\cite{DIC-00,VES-97,DIC-98,ALA-01}. Conversely,
criticality in sandpile models can be related to an underlying
depinning critical point \cite{TAN-88,VES-00}.  The scaling of the
power spectrum (PS) in sandpile models can be contrasted to 
avalanche induced crackling noise, which is typically
characterized by a power law distribution of amplitudes and by a 
non-trivial $1/f^\alpha$ spectrum \cite{SET-01}.  The most studied
condensed-matter examples 
include Barkhausen noise in ferromagnets \cite{DUR-05} and
acoustic emission in fracture \cite{PET-94,SAL-02} and plasticity
\cite{MIG-01}. 

In this letter, we show that, notwithstanding previous beliefs,
classical sandpile models display non-trivial $1/f^\alpha$ spectra. 
$\alpha < 2$ depends on the model and dimensions. 
We compute by numerical simulations the avalanche spectrum of two
classes of sandpile models: the original two dimensional BTW sandpile 
model and the stochastic two-state Manna model, in one, two and three 
dimensions (1d, 2d, 3d). These two models are now known to be in different 
universality classes. A further difference between the two classes of 
models is that stochastic sandpiles obey finite size scaling while the 
BTW model displays multiscaling \cite{stella}.  
We find that the power spectrum decays as $P(f) \sim f^{-\alpha}$, with
$\alpha=1.59 \pm 0.05$ for BTW and  
$\alpha = 1.44 \pm 0.05$, $\alpha=1.77 \pm 0.05$,
$\alpha=1.9 \pm 0.1$ for Manna in $d=1,2,3$, respectively.

The central idea as to why SOC models can exhibit varying $\alpha$,
with the details depending on the dimension and universality class,
is based on self-affine fractal dynamics. Consider the time series $V(t)$,
which records the number of ``topplings'' (local relaxation events) 
taking place in the sandpile during each parallel update of the whole 
lattice, one such update defining the unit of time. An avalanche is defined 
here as a connected sequence of non-zero values of $V(t)$. If the average size
(i.e. the total number of topplings) of such avalanches of duration $T$ 
scales as $\langle s(T) \rangle \sim T^{\gamma_{st}}$ and the dynamics 
is self-similar, then the average avalanche shape should follow 
\begin{equation} \label{eq:vshape}
V(T,t)=T^{\gamma_{st}-1}f_{shape}(t/T),
\label{shape}
\end{equation}
where $f_{shape}(x)$ is a scaling function \cite{KUN-00}. 
The form of the power spectrum is obtained averaging the energy
spectrum $E(f|s)$ of avalanches of size $s$, which in general
scales as
\begin{equation} \label{eq:energy}
E(f|s)=s^2 g_{E}(f^{\gamma_{st}}s),
\label{energy}
\end{equation}
where $g_{E}(x)$ is another scaling function \cite{KER-90,KUN-00}.
The power spectrum can then be written as $P(f)=\int
D(s)E(f|s)ds$, where $D(s) \sim s^{-\tau}$ is the probability 
distribution of avalanche sizes and the integral is bounded by the
upper cutoff $s^*$, so that  
\begin{equation}
P(f)= f^{-\gamma_{st}(3-\tau)} \int^{s^*f^{\gamma_{st}}}
dx x^{2-\tau} g_{E}(x). \label{eq:integral}
\end{equation}
If the integral in Eq.~(\ref{eq:integral}) is convergent, we obtain
$\alpha=\gamma_{st}(3-\tau)$ (as originally derived in Ref.~\cite{LIE-72}). 
In the opposite case, the final result
crucially depends on the asymptotic behavior of $g_{E}(x)$.  Kertesz
and Kiss assumed $g_{E}(x)\propto 1/(1+x^{2/\gamma_{st}})$, obtaining
$\alpha=2$ \cite{KER-90}. Jensen et al.~approximate the avalanche
shape with a box function, which implies $g_{E}(x)\propto
(1-\cos(x^{1/\gamma_{st}}))/x^{2/\gamma_{st}}$, yielding again
$\alpha=2$ \cite{JEN-89}. More recently, Kuntz and Sethna \cite{KUN-00} noticed that
if the toppling dynamics in the avalanche is a local process, the
released energy is an extensive function of the size $s$, or $E(f|s)
\sim s$. From Eq.~(\ref{energy}) it thus follows that $g_E(x) \sim
A/x$. This implies that for $\tau < 2$ (which is the case for sandpile
models), the integral in Eq.~(\ref{eq:integral}) is dominated by the
frequency dependent upper cutoff, yielding $\alpha=\gamma_{st}$.

Here we analyze numerically the above set of four test-cases.
We measure the shape of the pulse associated to an avalanche in
the models, the scaling behavior of the avalanche size for a given duration
and compute the power spectra. 
Sandpile models are defined on a $d-$dimensional hypercubic lattice. On
each site $i$ of the lattice the height is an integer variable $z_i$.
At each step the system is driven, a grain is dropped  on a randomly
chosen site raising its height by one unit ($z_i\to z_i+1$). 
When one of the sites reaches or exceeds
a threshold $z_c$ a ``toppling'' occurs: $z_i=z_i-z_c$ and
$z_j=z_j+1$, where $j$ represents the nearest neighbor sites of site
$i$. In the BTW model $z_c=2d$ and each nearest neighbor receives a
grain after the toppling of the site $i$. In the Manna model $z_c=2$
and therefore only two randomly chosen neighboring sites receive a
grain. A toppling can induce nearest-neighbor sites to topple on their
turn and so on, until all the lattice sites are below the critical
threshold.  This process defines an avalanche. We use parallel dynamics,
meaning that every overcritical site topples when the lattice is updated,
one such update defining the unit of time. The slow driving condition
implies that grains are added only when all the sites are
below the threshold. Grains can leave the system from the open boundaries.
After a transient, sandpile models reach a steady state with avalanches
of all sizes. Here we consider linear system sizes ranging from $L=1024$ to
$L=16384$ in 1d, from $L=64$ to $L=2048$ in 2d
and $L=32$ to $L=256$ in 3d.

\begin{figure}[htb]
\centerline{\psfig{file=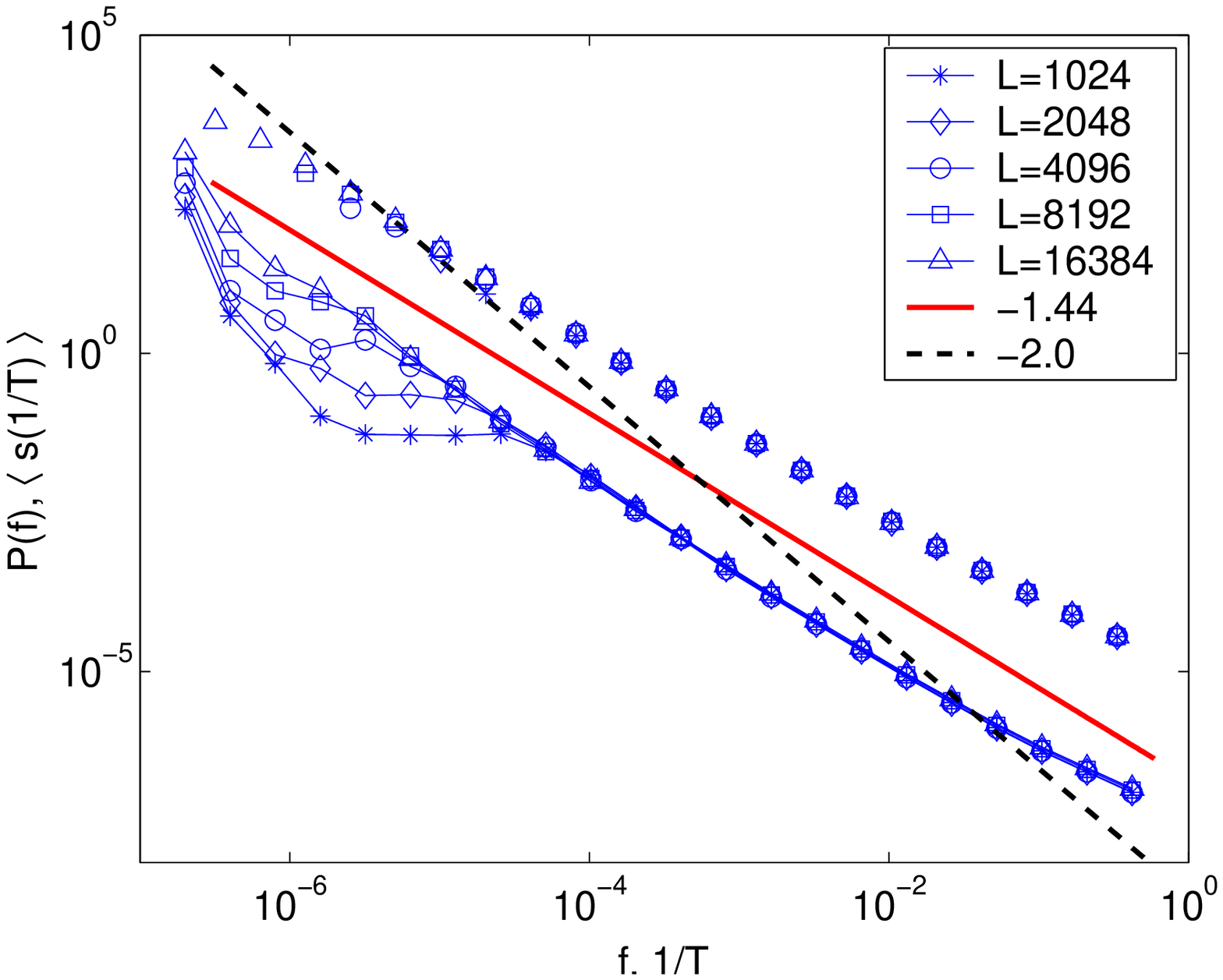,width=8cm}}
\centerline{\psfig{file=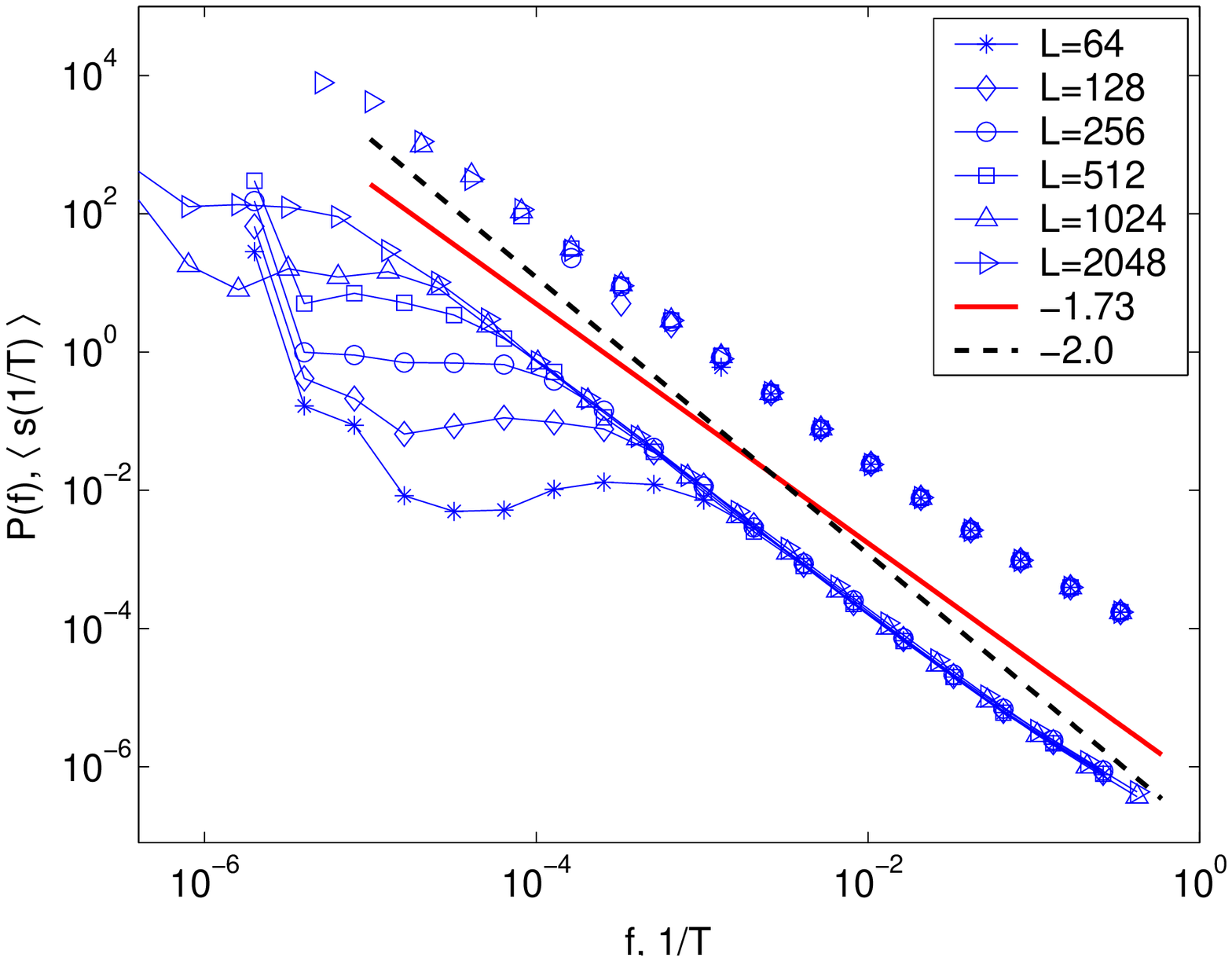,width=8cm}}
\centerline{\psfig{file=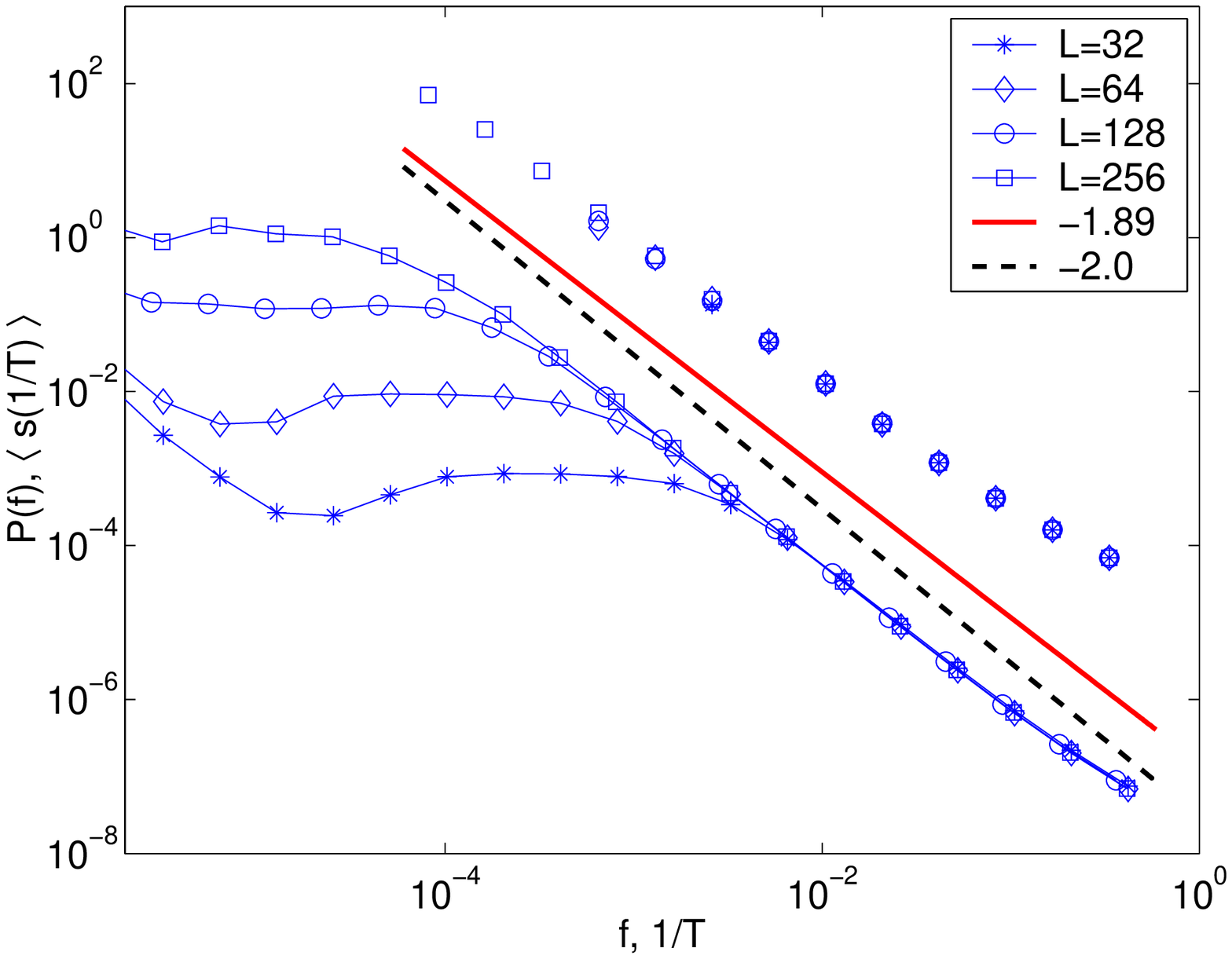,width=8cm}}
\caption{The scaling of the power spectra (symbols connected by a line) and 
$\langle s(1/T) \rangle$ (symbols without a connecting line)
of the Manna sandpile model for different system sizes,
Manna 1d (top), 2d (middle) and 3d (bottom).  
In all three cases, the slope of the spectrum is significantly 
different from $2$ (dashed line).}
\label{fig:1}
\end{figure}

\begin{figure}[htb]
\centerline{\psfig{file=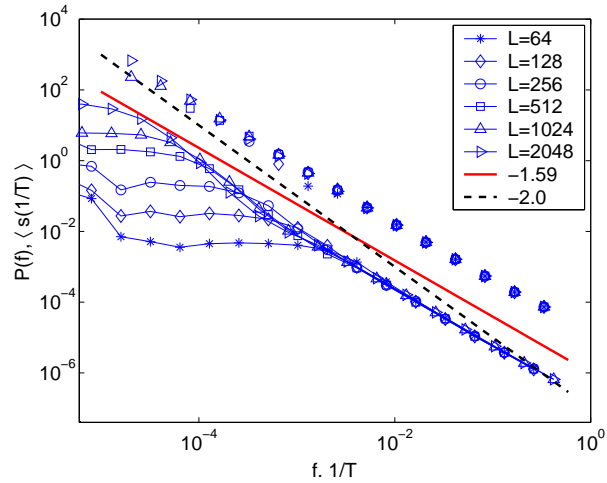,width=8cm}}
\caption{The scaling of the power spectrum (symbols connected by a line) and
$\langle s(1/T) \rangle$ (symbols without a connecting line)
of the BTW model in 2d for different system sizes. The slope of the high
frequency part of the power spectrum is again
significantly different from $2$ (dashed line).}
\label{fig:2}
\end{figure}

Power spectra are measured considering the signal produced by the
number of toppling events $V(t)$ as a function of time. Waiting times
between avalanches have no effect on the scaling of the high frequency
parts of the power spectra reflecting avalanche spectral properties
(i.e. frequencies corresponding to time scales smaller than that of
the duration of the longest avalanche). We have checked this e.g. by
inserting a constant number of zeros between every successive pair of
avalanches, and by means of a slow continuous drive, corresponding to
a  Poisson distribution of waiting times. Therefore in what follows, we simply
join the avalanches one after the other.  In Figs.~\ref{fig:1} and
~\ref{fig:2} we display the power spectra $P(f)$ and $\langle
s(1/T)\rangle$ of the Manna and BTW models for different system
sizes. While for the smallest lattices the power spectrum might be
fitted by a Lorentzian, for larger system sizes the tails are
definitely not scaling as $1/f^2$. Instead, by fitting to the scaling
parts of the power spectra (frequencies higher than the one
corresponding to the duration of the longest avalanche and lower than
inverse of a cross-over time after which the avalanches will have a
self-similar structure), we find $\alpha=1.59 \pm 0.05$ for BTW and
$\alpha=1.44 \pm 0.05$, $\alpha=1.77 \pm 0.05$ and $\alpha=1.9 \pm
0.1$ for Manna in 1d, 2d and 3d, respectively. Note that the results
are contrary to those in \cite{JEN-89,KER-90}, whose results were
obscured by the small system sizes reachable at the epoch.  Instead, at
least for the Manna model, the scaling of the power spectra follows
quite nicely that of $\langle s(1/T)\rangle \sim
(1/T)^{-\gamma_{st}}$, with $\gamma_{st}=1.44 \pm 0.05$,
$\gamma_{st}=1.73 \pm 0.05$ and $\gamma_{st}=1.9 \pm 0.1$ in 1d, 2d
and 3d, respectively \cite{LUB-05}. The PS of the BTW has a ``bump''
for small frequencies (Fig.~\ref{fig:2}), which shifts to still
smaller ones with increasing $L$.  Furthermore, for BTW, $\langle
s(1/T)\rangle$ exhibits slight curvature even for large avalanches,
but still the agreement is fair.

\begin{figure}[htb]
\centerline{\psfig{file=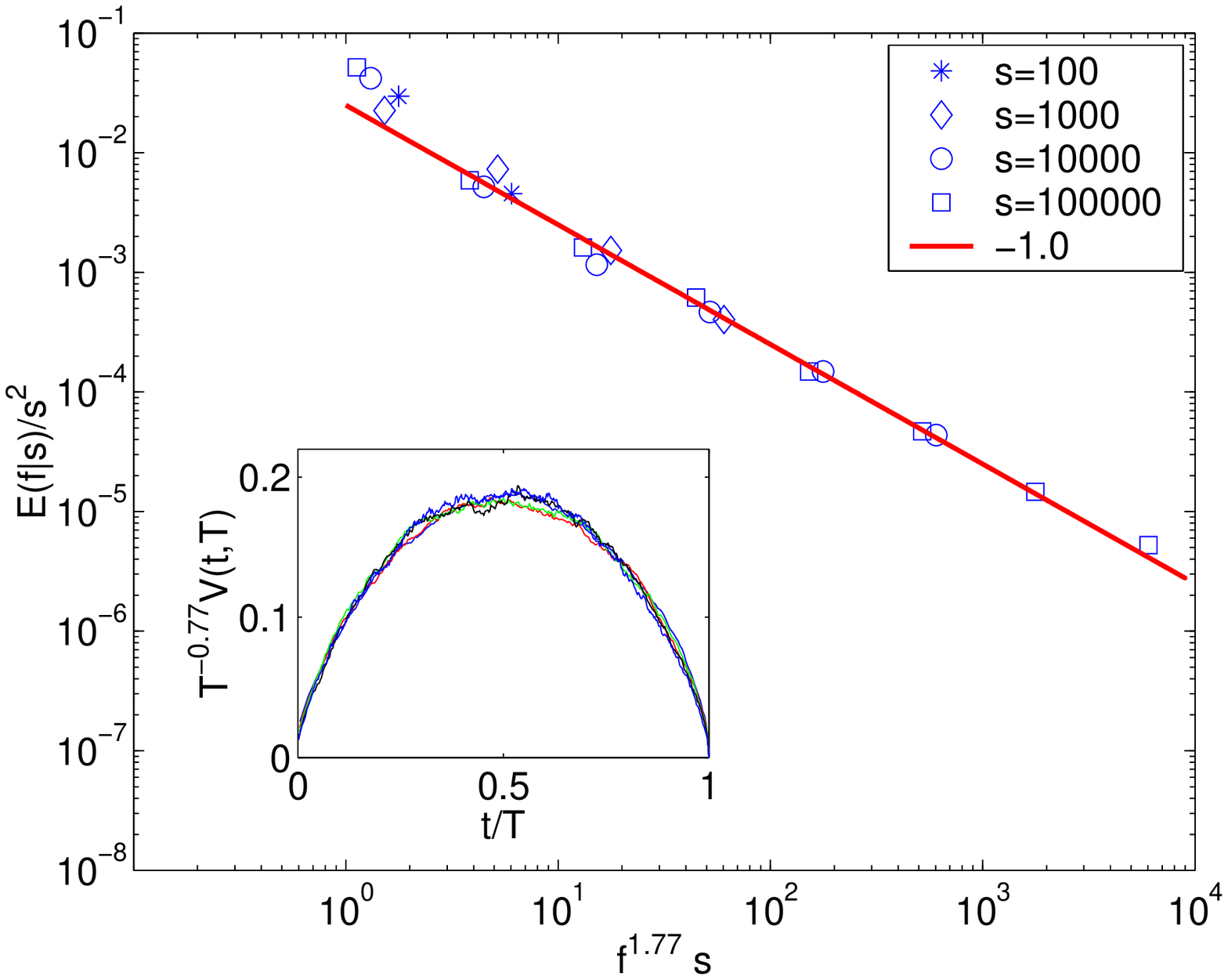,width=8cm}}
\caption{Main figure: the energy spectrum for different sizes $s$ is
collapsed according to Eq.~\protect\ref{energy}, and the scaling
function decays as $1/x$.  Inset: the collapse of the average
avalanche shape , avalanche durations ranging
from $T=200$ to $T=500$. All data are for the 2d Manna model.}
\label{fig:3}
\end{figure}

In order to check that the observed results follow directly
from the derivation outlined above, we compute the energy spectrum
$E(f|s)$. The results reported in Fig.~\ref{fig:3} confirm
the scaling behavior predicted by Eq.~\ref{energy} with $g_E \sim 1/x$.
For the avalanche shape, Eq.~(\ref{shape}), the models show slightly
different properties. Originally, Ref.~\cite{JEN-89} employed a simple
box function to approximate $f_{shape}$. This form is very far from
the correct one, as it is shown in the inset Fig.~\ref{fig:3}. While the
avalanche shape of the Manna model is symmetric, a more
detailed look at the BTW model reveals, that its avalanche shape {\em
can not be rescaled } as a function of duration $T$. 
The avalanches develop slowly an asymmetry, which
could be related with the observations of multiscaling in the BTW
model \cite{stella}.  For the stochastic Manna sandpiles, in 1d 
the assumption of scale-invariance holds nicely while in the 2d and 3d
cases relatively small avalanches show a cross-over behavior so that only
around $T \sim 100$ the scaling regime is reached. This naturally
implies corrections to scaling, but nevertheless $\alpha =
\gamma_{st}$ holds rather well (and in particular $\alpha < 2$).

The non-Lorenzian PS also persists if such systems are driven slowly
as long as the avalanches do not overlap much, corresponding to
timescales shorter than the inverse one related to the drive
frequency. It is theoretically important and intriguing that such a
relation can be established between the sandpile critical exponents
and the PS one. In spite of the relation between SOC and
non-equilibrium phase transitions, we still lack for analytical
predictions for the critical exponents in $d<4$, $d=4$ being the upper
critical dimension \cite{DIC-98}.  Thus, for $d\ge 4$ mean-field
exponents are valid, so that $\gamma_{st}=\alpha=2$. In two or three
dimensions, however, we would generally expect that $\alpha < 2$, and
thus in many real physical systems the expectation would be the same.

To summarize, self-organized criticality leads rather generally to
power-spectra that exhibit $1/f^\alpha$-noise with $\alpha < 2$. This
calls for, perhaps, a re-evaluation of experimental results in many
cases ranging from large systems met in solar and astro-physics to the
laboratory and carrying over to the understanding of brain dynamics
\cite{cogn}, biology and so forth. In other words, $\alpha<2$
does not imply the absence of SOC, but instead may indicate
exactly the contrary.  There are also
many theoretical issues that open up, including higher-order power
spectra \cite{PET-98}. In any case, we may safely conclude that the
central point of the original paper by Bak, Tang and Wiesenfeld
(i.e. the relation between avalanche scaling and non trivial power
spectrum) was ultimately correct, contrary to what has been believed to
be true from almost the beginning of the research on SOC and its
applications to various phenomena.

{\bf Acknowledgments}
We would like to thank the Center of Excellence program of
the Academy of Finland for financial support.

\References

\bibitem{BAK-87} 
        P. Bak, C. Tang, and K. Wiesenfeld, 
        Phys. Rev. Lett. {\bf 59}, 381 (1987); 
        Phys. Rev. A {\bf 38}, 364 (1988).
\bibitem{DUR-05}
        G. Durin and S. Zapperi,  in
        the Science of Hysteresis, Eds. G. Bertotti, I. Mayergoyz
        (Academic Press, New York, 2005), cond-mat/404512

\bibitem{FIE-95}
        S. Field, J. Witt, F. Nori, and X. Ling,
        Phys. Rev. Lett. {\bf 74}, 1206 (1995). 

\bibitem{PET-94}
        A. Petri, G. Paparo, A. Vespignani, 
        A. Alippi, and M. Costantini,
        Phys. Rev. Lett. {\bf 73}, 3423 (1994).
\bibitem{SAL-02}
        L.I. Salminen, A.I. Tolvanen, and M. J. Alava, 
        Phys. Rev. Lett. {\bf 89}, 185503 (2002).
\bibitem{reviews}
S. Chapman and N. Watkins, Space Sci. Rev. {\bf 95},  293 (2001);
T.Chang, S.W.Y. Tam, C.C. Wu, and O. Consolini,
 Space Sci. Rev. {\bf 107},  425 (2003).

\bibitem{journals}
The role of SOC in (laboratory) plasma, solar and magnetospheric physics 
has been discussed in several special issues of Phys.
Plasmas ({\bf 6:11}, (1999); {\bf 11:4}, 2004) and
see also J. Atmos. Sol.-Terr. Phys. {\bf 63: 13} 2001.
These contain numerous articles that refer to modeling
and empirical data, and also to power spectra.

\bibitem{magnetos}
T. Chang, Phys. Plasmas {\bf 6}, 4137 (1999);
V.M. Uritsky {\em et al.},
J. Geophys. Res. - Space Phys. {\bf 107}, 1426 (2002);
M.L. Sitnov  {\em et al.},
Phys. Rev. {\bf E65}, 016116 (2002).

\bibitem{bio}
T. Gisiger,
Biol. Rev. {\bf 76}, 161 (2001).

\bibitem{heart}
K. Kiyono {\em et al.},
        Phys. Rev. Lett. {\bf 93}, 178103 (2004).

\bibitem{cogn}
D.L. Gilden, 
Psychol. Rev. {\bf 108}, 33 (2001);
G.A. Worrell, S.D. Cranstoun, J. Echauz, and B. Litt, 
Neurorep. {\bf 13}, 2017 (2002);
J.M. Beggs and D. Plenz, J. Neurosci. {\bf 23}, 11167 (2003);
G. Buzsaki and A.  Draguhn, 
Science {\bf 304}, 1926 (2004);
E.J. Wagenmakers, S. Farrell, and R.  Ratcliff, 
Psych. Bull. Rev.
{\bf 11}, 579 (2004).

\bibitem{JEN-89}
H. J. Jensen, K. Christensen, and H. C. Fogedby,
Phys. Rev. B {\bf 40}, R7425 (1989).

\bibitem{KER-90}
J. Kertesz and L. B. Kiss,
J. Phys. A {\bf 23} L433 (1990).

\bibitem{1/f}
T. Hwa and M. Kardar, Phys. Rev. A  {\bf 45}, 7002 (1992);   
S. Maslov, C. Tang, and Y. C. Zhang, Phys. Rev. Lett. {\bf 83},  2449 (1999).

\bibitem{1/f2}
A. A. Ali, Phys. Rev. E {\bf 52}, R4595 (1995); J. Davidsen and M. Paczuski,
Phys. Rev. E {\bf 66}, 050101 (2002).

\bibitem{MAN-91}
        S. S. Manna, J. Phys. A {\bf 24}, L363 (1991).

\bibitem{DHA-99}
D. Dhar, Physica A {\bf 263}, 4 (1999) 
\bibitem{DIC-00} 
R. Dickman, M. A. Mu\~noz, A. Vespignani, and S. Zapperi, 
Braz. J. Phys. {\bf 30}, 27 (2000). 

\bibitem{VES-97}
        A. Vespignani and S. Zapperi,
        Phys. Rev. Lett. {\bf 78}, 4793 (1997);
        Phys. Rev. E {\bf 57}, 6345 (1998).

\bibitem{DIC-98}
        R. Dickman, A. Vespignani, and S. Zapperi,
        Phys. Rev. E {\bf 57}, 5095 (1998). 
        A. Vespignani, R. Dickman, M. A. Mu\~noz, and Stefano Zapperi,  
        Phys. Rev. Lett. {\bf 81}, 5676 (1998). 

\bibitem{ALA-01} 
M. Alava and K. B. Lauritsen, 
Europhys. Lett. {\bf 53}, 569 (2001)
 
\bibitem{ALA-04}
M. Alava, cond-mat/0307668, also in Advances in
{\em Condensed Matter and Statistical Mechanics}, Eds. E. Korutcheva
and R. Cuerno, (Nova Science Publishers, 2004.)

\bibitem{TAN-88} 
C. Tang and P. Bak,  
Phys. Rev. Lett. {\bf 60}, 2347 (1988). 

\bibitem{VES-00} 
A. Vespignani, R. Dickman, M. A. Mu\~noz, and S. Zapperi, 
Phys. Rev. E {\bf 62}, 4564 (2000). 

\bibitem{SET-01}
J.~Sethna, K.~A. Dahmen, and C.~R. Myers,  Nature {\bf 410},  242 (2001).

\bibitem{MIG-01}
        M. C. Miguel, A. Vespignani, S. Zapperi, J. Weiss, and J. R. Grasso, 
         Nature {\bf 410}, 667 (2001).

\bibitem{stella}
        M. De Menech, A. L. Stella, and C. Tebaldi,
        Phys. Rev. E {\bf 58}, R2677 (1998);
        C. Tebaldi, M. De Menech, and A. L. Stella,
        Phys. Rev. Lett. {\bf 83}, 3952 (1999).

\bibitem{KUN-00}
M.~C. Kuntz and J.~P. Sethna.
Phys. Rev. B {\bf 62}, 11699 (2000).

\bibitem{LIE-72}
U. Lieneweg and W. Grosse-Nobis
Inter. J. Magnetism \textbf{3}, 11 (1972).

\bibitem{LUB-05} One can derive from other
measured quantities approximative values for
$\gamma_{st}$ using published data (see e.g.~S.
L\"ubeck, cond-mat/0501250), and our
values are in reasonable agreement.

 \bibitem{PET-98}
J.~R. Petta, M.~B. Weissman, and G.~Durin.
 Phys. Rev. E {\bf 57}, 6363 (1998).

\endrefs
\end{document}